\documentclass[a4paper,12pt]{article}
\usepackage[latin2]{inputenc}
\usepackage[active]{srcltx}
\usepackage{german}
\usepackage{amsmath}
\usepackage{amsthm}
\usepackage{amsfonts}
\usepackage{amssymb}
\usepackage{textcomp}
\usepackage[bf]{caption}
\usepackage[sans]{dsfont}			
\usepackage{pifont}
\usepackage{color}
\usepackage{psfrag,graphicx}
\usepackage{hyperref}

\hypersetup{
	colorlinks = true,
	urlcolor = blue
}

\title{Domain-wall depinning assisted by pure spin currents}
\date{}
\author{D. Ilgaz$^1$, J. Nievendick$^1$, L. Heyne$^1$, D. Backes$^{1,2}$,\\ T. A. Moore$^1$, M. A. Nino$^3$, A. Locatelli$^3$, T. O. Mente\c{s}$^3$,\\A. v. Schmidsfeld$^1$, A. v. Bieren$^1$, J. Rhensius$^{1,2}$,\\ S. Krzyk$^1$, L. J. Heyderman$^2$ and M. Kläui$^1$\\ \begin{small}$^1$Fachbereich Physik, Universität Konstanz, Universitätsstraße 10, 78457 Konstanz\end{small}\\ \begin{small}$^2$ Laboratory for Micro- and Nanotechnology, Paul Scherrer Institut, CH-5232 Villigen PSI, Switzerland\end{small}\\ \begin{small}$^3$Sincrotrone Trieste S.C.p.A., Strada Statale 14, 34149 Basovizza, Trieste, ITALY\end{small}
}
\def \a {\alpha}
\def \b {\beta}
\def \g {\gamma}

\def \e {\epsilon}

\def \l {\lambda}
\def \m {\mu}

\def \r {\rho}

\def \D {\Delta}

\def \W {\Omega}
\def \T {\tfrac}
\def \R {\frac}

\def \H {\hbar}

\def \C {\partial}

\begin{document}
\maketitle
\abstract{
\begin{footnotesize}
We study the depinning of domain walls by pure diffusive spin currents in a nonlocal spin valve structure based on two ferromagnetic permalloy elements with copper as the nonmagnetic spin conduit. The injected spin current is absorbed by the second permalloy structure with a domain wall and from the dependence of the wall depinning field on the spin current density we find an efficiency of $\mathrm{6\cdot 10^{-14}T/(A/m^2)}$, which is more than an order of magnitude larger than for conventional current induced domain wall motion. Theoretically we reproduce this high efficiency, which arises from the surface torques exerted by the absorbed spin current that lead to efficient depinning.
\end{footnotesize}
}
\newpage
Spin currents and magnetoresistance effects have received much attention over the last two decades [1]. The reciprocal influence of spin polarized charge currents on the magnetization, which leads to switching in multilayer pillars [2] and the manipulation of magnetic domain walls (DWs) by currents, has become the focus of research due to the fundamental physics as well as possible applications [3, 4]. The manipulation of DWs has been attempted using spin polarized charge currents [5, 6, 7] or local Oersted fields [8]. However, DW depinning using Oersted fields is facing challenges regarding the scalability and for spin-polarized charge currents, the effective non-adiabatic torque is small for permalloy with wide walls ( $\b \ll$ 1) [9,10,11]. Furthermore, charge currents lead to Joule heating and at increased temperatures the spin torque efficiency further decreases [12].

A possible alternative approach is to employ pure spin currents, where the electrons diffuse without an associated net charge current. While the generation of spin currents involves energy dissipation, it can occur at a distant location from the device, which can thus be kept cool and still manipulated by the absorbed diffusive spin currents.

Nonlocal spin valves (NLSVs) are promising geometries to generate pure spin currents across transparent [13, 14, 15] or tunneling contacts [16, 17]. Recently, Yang \textit{et al.} have shown the reversal of the magnetic state of a permalloy disc in a NLSV geometry where the device could be used either in a nonlocal or a lateral spin valve contact setup, and the same critical current densities were observed since the same torques are acting [18]. This is expected to be radically different for the case of a domain wall in a ferromagnetic wire where the adiabatic and non-adiabatic torques exerted by a combined charge and spin polarized current flowing in the wire across the DW and the torques exerted by a spin current absorbed at the DW position will be fundamentally different but to date this has not been investigated.

We present in this paper depinning of DWs assisted by pure spin currents. We determine the spin diffusion length in Cu and the spin polarization in permalloy ($\mathrm{Ni_{80}Fe_{20}}$, Py), and from this we calculate the spin current that diffuses to the ferromagnetic structure where a DW is positioned. We measure the spin current assisted DW depinning and find a large efficiency of the spin current induced torques. This can be explained by the large interface torque that stems from the perpendicular orientation of the magnetization in the domain wall with respect to the spins in the spin current acting on the surface layers.

The two samples examined here (referred to as A and B hereafter) were fabricated in a two step lithography process [19] and a scanning electron microscope (SEM) image of such a sample is shown in Fig. 1 (a). In Fig 1 (b), an X-ray Magnetic Circular Dichroism Photoemission Electron Microscopy (XMCD-PEEM) [20] image of the magnetization configuration with a DW and a corresponding micromagnetic simulation [21] using the same geometry are presented (Fig. 1(c)). First the two Py elements shown in Fig. 1 (a) were deposited with a thickness of 17 nm and a width of 300 nm and 600 nm for the wire and the half ring wire respectively. Before the deposition of 50 nm copper (Cu) as the nonmagnetic material, ion milling was used to clean the interface. On top of the Cu layers, 2 nm of Au was deposited to prevent them from oxidation. The width of the central Cu wire is 330 nm for sample A and 490 nm for sample B. The edge-to-edge distance between the Py wires was 295 nm in sample A and 110 nm in sample B. We did not take advantage of tunnel barriers at the ferromagnetic-nonferromagnetic interfaces, since, although better injection efficiencies can be achieved [16, 17], this strongly limits the maximum charge current. Transport measurements were conducted in a cryostat at 4.2 K using a standard lock-in technique and an in-plane rotatable external magnetic field.

We measure the nonlocal spin signal on both samples (see Fig. 2) with a peak applied charge current density of $2\cdot10^{11}\T{A}{m^2}$ in the Py wire. The origin of these signals is a charge current $I_C$ driven from a ferromagnet (FM) into a nonferromagnet (NM), which generates a spin accumulation diffusing to a second ferromagnet [14]. From the calculations in [22, 23] for a nonlocal geometry with two different FMs, one obtains a nonlocal spin voltage of
\begin{flalign}
	\D V_{NL}(d) = \R{\a^2_F I_C R_{S,F1}R_{S,F2}R_{S,N}}{\exp \big(\R{d}{\l_{N}}\big) \big[R_{S,N}(R_{S,F1}+R_{S,F2})+2R_{S,F1}R_{S,F2}\big]   +R_{S,N}^2\sinh \big(\R{d}{\l_N}\big)}
\end{flalign}
at the second interface, which describes the difference between a parallel and antiparallel magnetic configuration. Here, $d$ is the distance between both FMs, $\a_F$ the spin polarization in the FM and $R_{S,i}$ is the spin resistance with $R_{S,i} = 2\r_i\l_i/(S(1-\a^2_i))$, where $\r_i$ is the resistivity, $\l_i$ the spin diffusion length of the specific material and $S$ the cross-sectional area. The spin current at the second interface is then given by $I_S(d) = \D V_{NL}(d)/(\a_FR_{S,F2})$, and the measured nonlocal spin resistance change is defined as $\D R_{NL} = \D V_{NL}/I_C$.

The jumps in the nonlocal spin resistance signal (Fig. 2) correspond to the switching of the FM wire and half ring (see sketch in Fig. 2). The spin signal increases from 20.9 $\m\W$ for an edge-to-edge distance of 295 nm in sample A to 88.8 $\m\W$ in sample B with a spacing of 110 nm between both Py wires. Note that the measured spin resistance signal is not symmetric around 0 $\W$ and this can originate from an inhomogeneous current distribution [24, 25]. Furthermore the switching fields are not exactly equal in both samples, which is probably due to slight variations in the geometry and edge roughness.

We used the following approach to determine the spin diffusion length in the Cu wire and the spin polarization of the Py stripes: Due to the small spin diffusion length in Py [26, 27], the cross-section areas for both ferromagnets are defined as the interface cross-section (width of the permalloy structure times width of the Cu structure). This results in a cubic dependence on the Cu width of the numerator and a square dependence of all terms in the denominator in Eq. 1. Multiplying Eq. 1 with the Cu width makes the right side thus independent of the Cu width, thus allowing us to use two samples with different Cu widths for the analysis. The resistivities used are 25 $\m\W$cm for the bulk Py value and 2.2 $\m\W$cm for the Cu respectively. To determine the spin diffusion length in Cu and the spin polarization of Py, we use a fixed value of 5 nm for the spin diffusion length in Py as this has been determined independently by two groups [26, 27]. By fitting these used values to the modified equation, we obtain a spin polarization of $\a_{FM} = 43\pm 1$\% and a spin diffusion length of $\l_N = \l_{Cu} = 134\pm 12$ nm. While the spin polarization found is in agreement with results of Soulen \textit{et al.} [28], the spin diffusion length is below the results of Ji \textit{et al.} [15] and those of Jedema \textit{et al.} [29].

Using these results, we can calculate the spin current that arrives at the Py halfring compared to the charge current injected between contacts 3 and 4 and we find a ratio of $\mathrm{I_S/I_C} = (1.2\pm 0.1)\cdot 10^{-2}$.

We now employ these spin currents to manipulate the magnetization. We study their influence on the depinning behaviour of a transverse DW (TDW) in the half ring in sample B. The TDW was nucleated with a field as described in [30], with a resulting magnetic structure and a corresponding OOMMF simulation [21] shown in Fig. 1 (b) and (c). The TDW is positioned below the central Cu wire slightly off center to the right (see sketch in Fig. 3 (d)). During the experiment a spin current is generated by a 50 $\m$s long charge pulse between contacts 3 and 4, which is then absorbed by the TDW in the Py half ring. The position and ultimately the depinning of the TDW is determined by the voltage drop due to the AMR signal between contacts 8 and 9 when applying a small ac lock-in current ($10^{10} A/m^2$) between 3 and 10 [30].

The dependence of the depinning field as a function of the current amplitude is now shown in Fig. 3 (a): For negative currents, the depinning field decreases with increasing current amplitude. For positive currents, where the spin current and the applied field act in opposite directions, one would expect an increase in the depinning field if DC currents are used. But as we use current pulses, the wall depins in between pulses at a field that corresponds to the zero current depinning field, a behavior that we have discussed previously in [12]. The constant depinning field for positive currents also shows that there was no significant Joule heating affecting the depinning. In order to compare our findings with the results of current induced domain wall motion (CIDM), where the combined spin and charge current flows in the ferromagnet, we divide the spin current by the cross-sectional area of the Py half ring, which results in the spin current densities shown on the upper x-axis of Fig. 3 (a). We obtain a depinning efficiency of $\mathrm{(6\pm 1)\cdot 10^{-14}\;Tm^2/A}$ and by extrapolation a spin current density of $\mathrm{(7\pm 2)\cdot 10^{10}\;A/m^2}$ at which the DW would depin without any external field. Compared with CIDM, the efficiency is larger by an order of magnitude ($\e_{\mathrm{CIDM}} \approx 5\cdot 10^{-15}\;Tm^2/A$ [12]) and the extrapolated required current density for depinning ($j_{\mathrm{CIDM}}\approx 2\cdot 10^{12}\;A/m^2$ [12]) is about thirty times smaller.

To demonstrate that it is the spin current that acts on the DW, we simulate the influence of Oersted fields created by the pulses between contacts 3 and 4 with the maximum charge current that was applied in our experiments. We find a maximum field of less than 10 Oe at the edge and an average field of 1 Oe in the area of the DW, which is negligible compared to the depinning field. Furthermore, we have repeated the experiment with the DW at distances of a few hundred nm from the central Cu wire and we see that in this case the depinning of a DW is not affected by the currents, which excludes Oersted field effects and points to spin currents effects.

To theoretically explain the observed high efficiency, we look at the fundamental differences between the torques caused by spin polarized charge currents flowing in a single Py wire and the lateral spin valve geometry used here. The change of the magnetization $\vec m$ (here a dimensionless unit vector) in the case of a spin current being absorbed at a FM-NM interface is given by [31]
\begin{flalign}
	\R{\C \vec{m}}{\C t} = \R{\g\H}{2eM_SV} \vec m \times \vec I_S \times \vec m.
\end{flalign}
Here, $\H$ is the Planck constant, $M_S$ the saturation magnetization of the FM, $\g = \T{2\m_B}{\H}$ the gyromagnetic ratio ($\m_B$ being the Bohr magneton), $\vec I_S$ the orientation of the spin current injected into the FM and $V$ the volume affected by the noncollinear torque. For our case, $V$ is defined by the penetration depth of the spin currents times the surface where the spin currents enter and act on the FM. This surface is determined in one direction by the DW width and in the other by the wire thickness (assuming all the spins enter by the side wall) or the wire width plus the thickness (assuming that the spins enter by the side and the top surface of the Py half ring). In our experiment, the orientation in the spin current ($\vec I_S$) is perpendicular to the magnetization ($\vec m$) inside the TDW (see Fig. 3 (d)) leading to a maximized torque with a magnitude of
\begin{flalign}
	\bigg|\R{\C \vec{m}}{\C t}\bigg|_{SC} =\R{I_S\m_B}{eM_SV}.
\end{flalign}
This has to be compared with the spin polarized charge current induced magnetization change. Assuming $\a = \b$ [9, 10, 11], the magnitude of the torque for a DW along the x-axis is given by [32]:
\begin{flalign}
	\bigg|\R{\C \vec{m}}{\C t}\bigg|_{CIDM} = \R{\a_FI_C\m_B}{AeM_S}\Bigg|\R{\C \vec{m}}{\C x}\Bigg|,
\end{flalign}
Here, $I_C$ is the charge current sent through a FM wire with a cross sectional area $A$. Dividing Eq. (3) by Eq. (4), we obtain:
\begin{flalign}
	\R{|\C \vec{m}/\C t|}{|\C \vec{m}/\C t|_{CIDM}} = \R{I_SA}{\a_F I_CV(\C m /\C x)}
\end{flalign}
Depending on the volume affected by the spin currents, the ratio becomes $\simeq$ 25 (assuming the side and the top of the FM) to $\simeq$ 900 (assuming only the side). The magnetization gradient $\R{\C m}{\C x}$ is given by the domain wall width $l_{DW}$: $\R{\C m}{\C x} = \R{2}{l_{DW}}$ and the magnetic dephasing length used is 0.8 nm [33]. Thus, these calculations show that for pure spin current induced DW depinning the impact on the magnetic moments at the edge of the FM wire are one to two orders of magnitude larger than for CIDM. For DWs pinned by the edge, this gives an estimate of the increased depinning efficiency in line with our measurements.

In conclusion, we have shown that the depinning of DWs can be efficiently assisted by nonlocal spin currents  due to the large torque that then acts on the surface layers of the FM, where the domain wall pinning originates.

The authors acknowledge support by the DFG (SFB 767, KL1811), the ERC Starting Independent Researcher Grant (ERC-2007-Stg 208162), the EU RTN Spinswitch (MRTN-CT-2006-035327), as well as the Samsung Advanced Institute of Technology and would like to thank A. Brataas for fruitful discussions.
\subsection*{References}
\begin{footnotesize}
[1] M. N. Baibich \textit{et al.}, \href{http://dx.doi.org/10.1103/PhysRevLett.61.2472}{Phys. Rev. Lett.} \textbf{61}, 2472 (1988), G. Binasch \textit{et al.}, \href{http://dx.doi.org/10.1103/PhysRevB.39.4828}{Phys. Rev. B} \textbf{39}, 4828 (1989)\newline
[2] J.A. Katine \textit{et al.}, \href{http://dx.doi.org/10.1103/PhysRevLett.84.3149}{Phys. Rev. Lett.} \textbf{84}, 3129 (2000)\newline
[3] R. P. Cowburn \textit{et al.}, Patent WO/2007/132174 (2007)\newline 
[4] S. S. P. Parkin, M. Hayashi and L. Thomas, \href{http://dx.doi.org/10.1126/science.1145799}{Science} \textbf{320}, 190 (2008)\newline
[5] G. Tatara, H. Kohno and J. Shibata, \href{http://dx.doi.org/10.1016/j.physrep.2008.07.003}{Phys. Rep.} \textbf{468}, 213 (2008)\newline
[6] A. Yamaguchi \textit{et al.}, \href{http://dx.doi.org/10.1103/PhysRevLett.92.077205}{Phys. Rev. Lett.} \textbf{92}, 077205 (2004)\newline
[7] M. Kläui \textit{et al.}, \href{http://dx.doi.org/10.1103/PhysRevLett.94.106601}{Phys. Rev. Lett.} \textbf{94}, 106601 (2005)\newline
[8] D. Ilgaz \textit{et al.}, \href{http://dx.doi.org/10.1063/1.2990629}{Appl. Phys. Lett.} \textbf{93}, 132503 (2008)\newline
[9] R. Moriya \textit{et al.}, \href{http://dx.doi.org/10.1038/nphys936}{Nat. Phys.} \textbf{4}, 368 (2008)\newline
[10] L. Thomas \textit{et al.}, \href{http://dx.doi.org/10.1038/nature05093}{Nature} \textbf{443}, 197 (2006)\newline
[11] S. Lepadatu \textit{et al.}, \href{http://dx.doi.org/10.1103/PhysRevB.81.020413}{Phys. Rev. B} \textbf{81}, 020413 (2010) \newline
[12] M. Laufenberg \textit{et al.}, \href{http://dx.doi.org/10.1103/PhysRevLett.97.046602}{Phys. Rev. Lett.} \textbf{97}, 046602 (2006)\newline
[13] M. Johnson and R. H. Silsbee, \href{http://dx.doi.org/10.1103/PhysRevLett.55.1790}{Phys. Rev. Lett.} \textbf{55}, 1790 (1985) \newline
[14] F. J. Jedema, A. T. Filip and B. J. van Wees, \href{http://dx.doi.org/10.1038/35066533}{Nature} \textbf{410}, 345 (2001)\newline
[15] Y. Ji \textit{et al.}, \href{http://dx.doi.org/10.1063/1.2170138}{Appl. Phys. Lett.} \textbf{88}, 052509 (2006)\newline
[16] S. O. Valenzuela and M. Tinkham, \href{http://dx.doi.org/10.1063/1.1830685}{Appl. Phys. Lett.} \textbf{85}, 5914 (2004)\newline
[17] A. Vogel et al., \href{http://dx.doi.org/10.1063/1.3109787}{Appl. Phys. Lett.} \textbf{94}, 122510 (2009)\newline
[18] T. Yang, T. Kimura and Y. Otani, \href{http://dx.doi.org/10.1038/nphys1095}{Nat. Phys.} \textbf{4}, 851 (2008)\newline
[19] L. J. Heyderman \textit{et al.} \href{http://dx.doi.org/10.1016/j.mee.2004.03.051}{Microelectr. Eng.} \textbf{73-74}, 780 (2004)\newline
[20] J. Stöhr \textit{et al.}, Science \textbf{258}, 658 (1993)\newline
[21] \href{http://math.nist.gov/oommf/}{http://math.nist.gov/oommf/}, cell size 5x5x17nm, material: Py ($M_S=860\T{kA}{m},$ $A=1.3\cdot10^{-11}\T{J}{m}$)\newline
[22] S. Takahashi and S. Maekawa, \href{http://dx.doi.org/10.1103/PhysRevB.67.052409}{Phys. Rev. B} \textbf{67}, 052409 (2003), J. Fabian and I. \v{Z}uti\'{c} in \textit{Spintronics - From GMR to Quantum Information, Lecture Notes 40th Spring School}, edited by S. Blügel, D. Bürgler, M. Morgenstern, C. M. Schneider and R. Waser (Forschungszentrum Jülich, Jülich, 2009)\newline
[23] T. Kimura, Y. Otani and J. Hamrle, \href{http://dx.doi.org/10.1103/PhysRevB.73.132405}{Phys. Rev. B} \textbf{73}, 132405 (2006)\newline
[24] F. Casanova \textit{et al.}, \href{http://dx.doi.org/10.1103/PhysRevB.79.184415}{Phys. Rev. B} \textbf{79}, 184415 (2009)\newline
[25] M. Johnson and R. H. Silsbee, \href{http://dx.doi.org/10.1103/PhysRevB.76.153107}{Phys. Rev. B} \textbf{76}, 153107 (2007)\newline
[26] S. Dubois \textit{et al.}, \href{http://dx.doi.org/10.1103/PhysRevB.60.477}{Phys. Rev. B} \textbf{60}, 477 (1999)\newline
[27] S. D. Steenwyk \textit{et al.}, \href{http://dx.doi.org/10.1016/S0304-8853(97)00061-9}{J. Magn. Magn. Mater.} \textbf{170}, L1 (1997)\newline
[28] R. J. Soulen, Jr. \textit{et al.}, \href{http://dx.doi.org/10.1126/science.282.5386.85}{Science} \textbf{282}, 85 (1998)\newline
[29] F. J. Jedema \textit{et al.}, \href{http://dx.doi.org/10.1103/PhysRevB.67.085319}{Phys. Rev. B} \textbf{67}, 085319 (2003)\newline
[30] M. Kläui \textit{et al.}, \href{http://dx.doi.org/10.1103/PhysRevLett.90.097202}{Phys. Rev. Lett.} \textbf{90}, 097202 (2003)\newline
[31] Y. Tserkovnyak \textit{et al.}, \href{http://dx.doi.org/10.1103/RevModPhys.77.1375}{Rev. Mod. Phys.} \textbf{77}, 1375 (2005)\newline
[32] Z. Li and S. Zhang, \href{http://dx.doi.org/10.1103/PhysRevLett.92.207203}{Phys. Rev. Lett.} \textbf{92}, 207203 (2004)\newline
[33] S. Urazhdin, R. Loloee and W. P. Pratt, Jr., \href{http://dx.doi.org/10.1103/PhysRevB.71.100401}{Phys. Rev. B} \textbf{71}, 100401 (2005)
\end{footnotesize}\newpage
\begin{center}
	\includegraphics[width=0.48\textwidth]{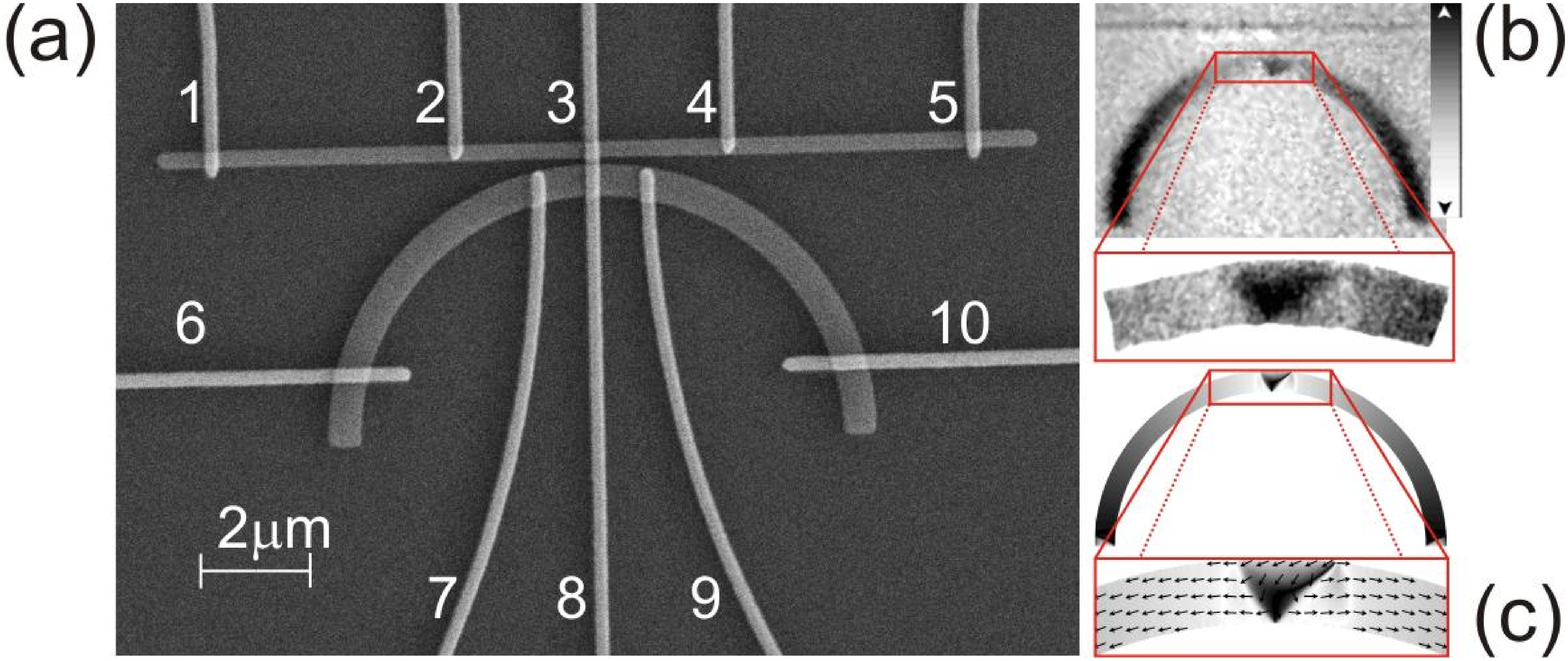}
\end{center}
\captionof*{figure}{\textbf{Fig. 1: }(a) Scanning electron microscopy (SEM) image of the nonlocal spin valve geometry of the samples used in the experiment with the contacts numbered 1 to 10. The bright stripes represent the Cu contacts, while the darker stripes constitute the Py wire and half ring. (b) XMCD-PEEM image of the spin configuration and the enlarged image of the transverse domain wall prior to contacting. The shades of grey visualize the vertical component of the magnetization configurations, which is in agreement with the micromagnetic simulation shown in (c).}\newpage
\begin{center}
	\includegraphics[width=0.48\textwidth]{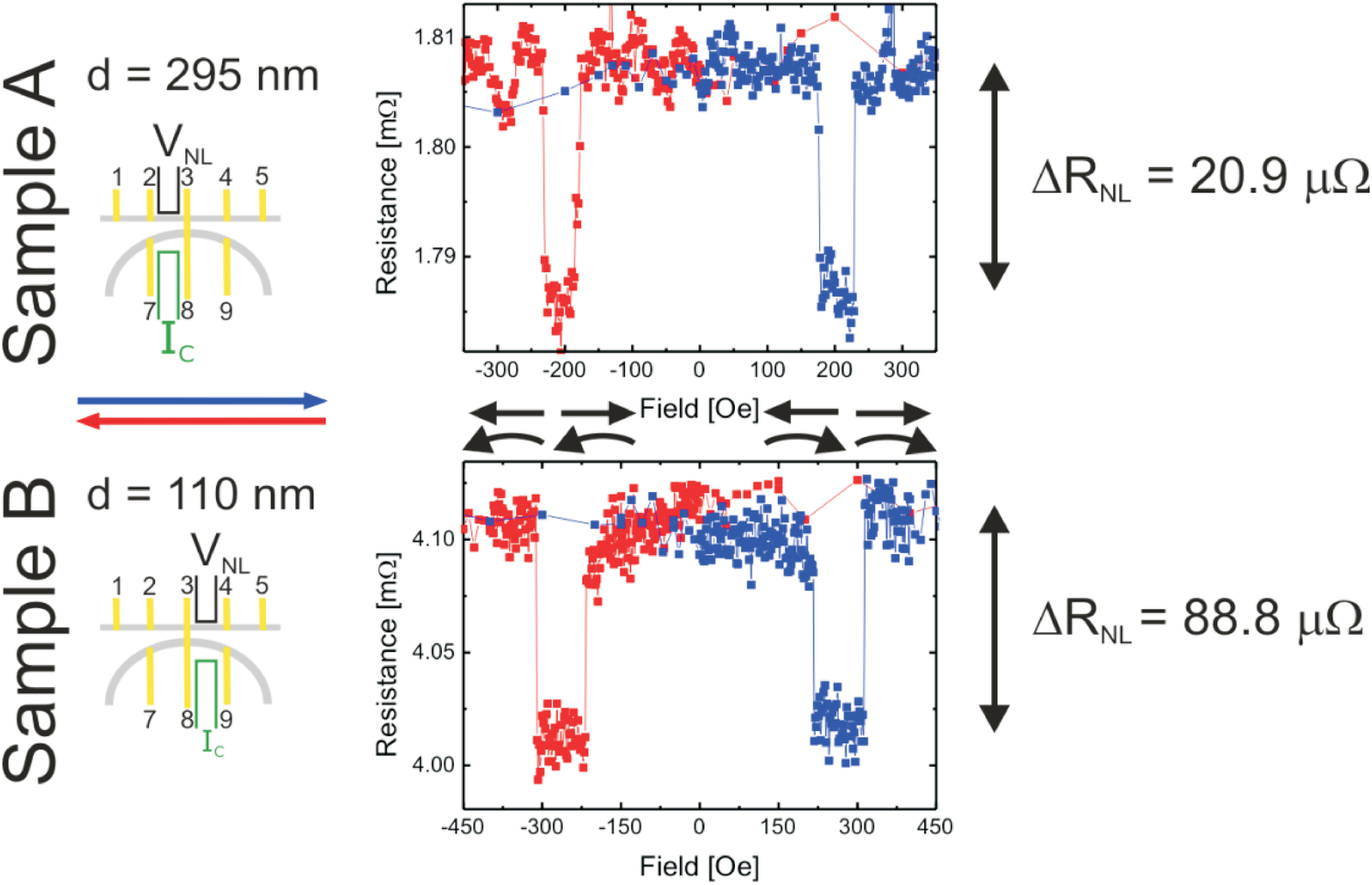}
\end{center}
\captionof*{figure}{\textbf{Fig. 2: }Non-local spin resistance measurements ($ R_{NL} \equiv \T{V_{NL}}{I_C}$) for both samples with a depiction of the corresponding current and voltage contact setup sketched to the left of the diagrams and the respective resistance differences to the right. For sample B with a smaller edge-to-edge distance d between the injector- and detector, the nonlocal spin signal increases to $88.8\; \m\W$. The arrows between the diagrams indicate the magnetic orientation of the FM wire and halfring.}\newpage

\begin{center}
	\includegraphics[width=0.6\textwidth]{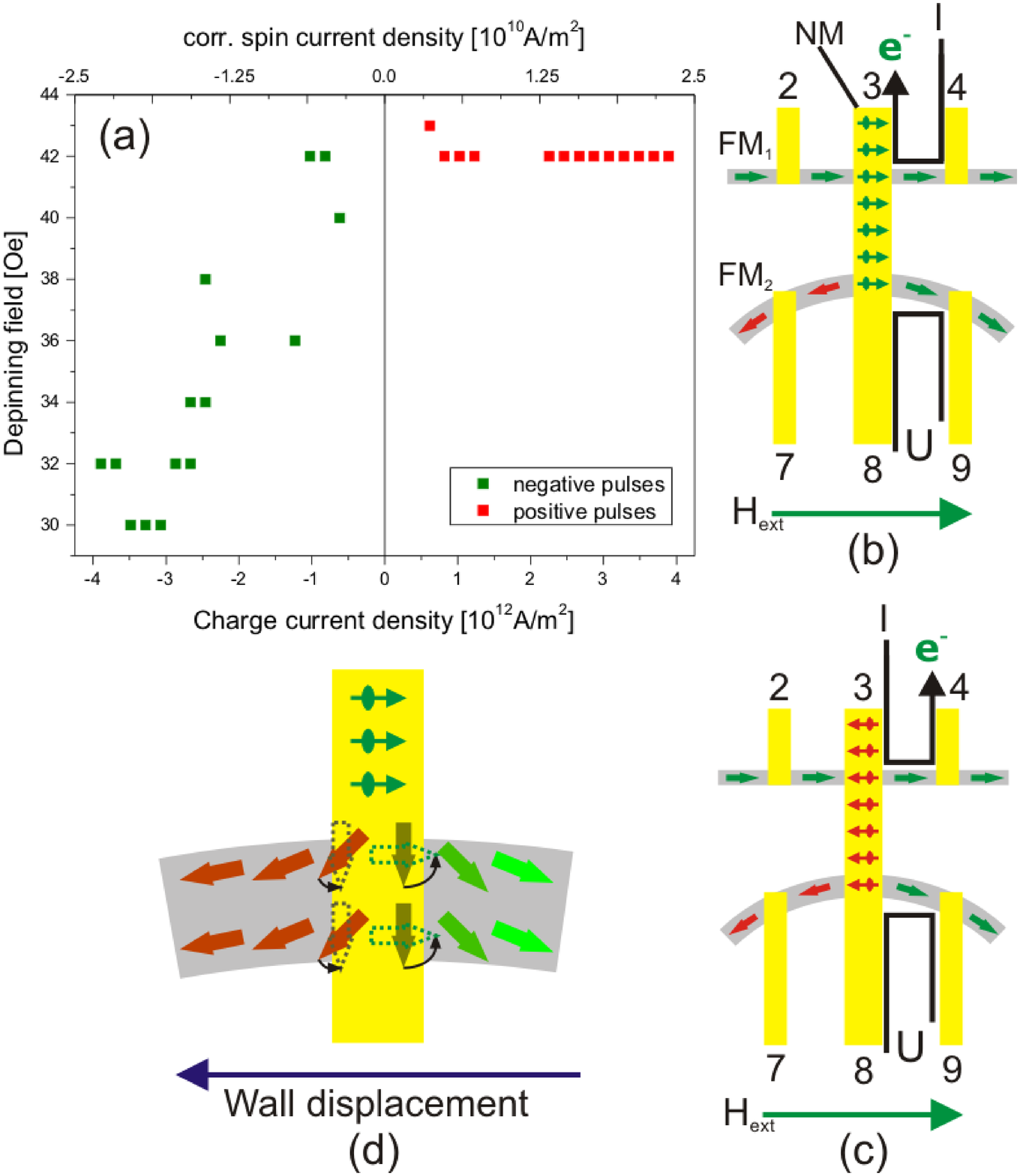}
\end{center}
\captionof*{figure}{\textbf{Fig. 3: }(a) Depinning fields as a function of the applied charge currents and the resulting corresponding spin current density (top x-axis). The direction of the charge current flow I and the injected pure diffusive spin current (arrows with circles on the central yellow NM spin conduit) are shown in (b) for negative current pulses. The situation for positive pulses is depicted in (c). If the electron (charge) current flows from the ferromagnet (FM$_1$) into the non-ferromagnet (NM) as in (b), the spins in the spin current (green arrows with circles) are oriented parallel to the magnetization of the ferromagnet (FM$_1$). In the opposite case (c) the spins in the spin current (red arrows with circles) are oriented antiparallel to the magnetic orientation of FM$_1$. In (d) we show for negative current pulses, the torque (Eq. (2)) exerted by the spin currents that is absorbed from the NM into the FM$_2$ where the transverse DW is located. The spin current absorption leads to a rotation of the original magnetization (large arrows) in FM$_2$ below the NM wire counterclockwise (indicated by the black arrows in (d)). The resulting magnetization direction is shown by the dotted large arrows, meaning that the TDW is effectively displaced to the left.}
\end{document}